\documentclass[prl,aps,showpacs,twocolumn,floatfix]
{revtex4-1}
\usepackage{graphicx}
\usepackage{bm}
\usepackage{epsfig}
\usepackage{amsmath}
\usepackage{amssymb}

\newcommand{\CA}{\cal A}
\newcommand{\C}{\cal C}

\newcommand{\al}{\alpha}

\newcommand{\gm}{\gamma}

\newcommand{\sg}{\sigma}

\newcommand{\lm}{\lambda}

\renewcommand{\th}{\theta}
\newcommand{\eq}{\begin{eqnarray}}
\newcommand{\eqx}{\end{eqnarray}}
\newcommand{\bal}{\begin{align}}
\newcommand{\eal}{\end{align}}
\newcommand{\ba}{\begin{equation}}
\newcommand{\ea}{\end{equation}}
\newcommand{\fr}{\frac}

\newcommand\n{\nonumber \\}

\begin{document}
\vspace{1cm}
\title{The scaling Pomeron}

{\it In honor of Andrzej Bialas for his $90^{th}$ anniversary}

\author{R. Peschanski}
\email{robi.peschanski@ipht.fr}

\author{B. G. Giraud}
\email{bertrand.giraud@ipht.fr}

\affiliation{Institut de Physique Th\'eorique,
Centre d'Etudes Saclay, 91191 Gif-sur-Yvette, France}

\begin{abstract}
We examine the Regge theoretical properties  for the scaling observed in pp elastic scattering differential cross-sections at the LHC. A positive signature amplitude  (i.e. the Pomeron) with scaling properties has been derived. It is found to describe the dip-bump region of momentum transfer at LHC energies in agreement with data. We derive the analytic continuation  in the whole plane of  the t-channel partial waves of index $l_t$  specific to the  Regge formalism. The analytic form  of the amplitude exhibits a specific scaling property without singularities, except for a series of poles in the  $l_t$ real axis at fractional values.
\end{abstract}
\date{\today}
\maketitle

\subsection{1. Scaling proton-proton scattering amplitudes}
\label{1}

Recently \cite{scaling}, scaling properties of proton-proton 
elastic scattering at LHC energies have been revealed, based on 
the elastic differential cross-sections measured by the TOTEM 
collaboration \cite{TOTEM}. 
Consider the  $pp$  differential cross-section  
\eq
   {\fr {d\sg}{dt}}^{pp\!}(s,t)\ &=& \ {1  \over  (s/ { {\rm TeV^2}})^2}\ \vert{\CA}^{pp}(s,t)\vert^2\ ,
\label{proton}
\eqx
where the usual Mandelstam
variables are $s,$  the positive center-of-mass energy 
squared and $t,$ the negative of the momentum transfer squared.
${\CA}^{pp}(s,t)$ is the proton-proton elastic  
spin-averaged scattering amplitude. For 
convenience, formula Eq.\eqref
{proton} is normalized in a dimensionless  way such that the 
amplitude  is dimensioned to an energy squared. 

 The scaling 
 properties empirically found in \cite{scaling} are such that a 
conveniently scaled cross section
 appears, within error bars, to depend only on one variable $t^{**}(s,t)= (s/ {\rm TeV^2})^{.065 }|t|^{.72} $. One finds
\ba
(s/ { {\rm TeV^2}})^{-.305}  {\fr {d\sg^{pp}}{dt}}(s,t) =  
f \left(t^{**}\right)\ ,
\label{crossscalingzero}
\ea
where $t$ is measured in $-{\rm GeV^2}$. The empirical scaling property, Eq.\eqref{crossscalingzero}, means that 
a model proton-proton elastic 
amplitude ${\CA}^{pp}(s,t)$ in Eq.\eqref{proton} may be written \cite{scaling} using one scaling variable 
\ba
\tau\ \equiv \ t\times {(s/ { {\rm TeV^2}})^{\gm/2} } \ ,
 \label{scalingvariable}
\ea
and two scaling exponents, $\al=0.35$ and $\gm= 
{.13}/{.72}=0.1806,$ namely
\ba
{\CA}^{pp}(s,t) =  (s/ { {\rm TeV^2}})^{\al/2}\ 
 F(\tau)\ .
\label{crossscaling}
\ea 
Indeed, in Ref.\cite{scaling}, one  finds a concrete realization of the 
amplitude ${\CA}^{pp}$ verifying the scaling Eq.\eqref
{crossscalingzero} when performing a fit of the measured cross-sections 
measured 
by the TOTEM collaboration \cite{TOTEM} in the dip-bump region of 
transverse momentum,  namely,
\eq
{\cal A}^{pp} &=& \ e^{i\th} 
({\cal A}_{1}-e^{i\varphi} {\cal A}_{2})\n
{\cal A}_{j=1,2} &=& \ i \ N_j^0 \ 
(s/ { {\rm TeV^2}})^{\fr \al 2 +1} \ e^{B_j^0 \tau }                                                  \ ,
\label{amplifit}
\eqx
where the parameters $N_j^0,B_j^0,  {j=1,2} $ and the complex phase
 $\varphi$  were phenomenologically adjusted \cite{scaling} 
by the fit. Then the overall phase $\th$ was fixed by the known ratio of the real over imaginary 
part of the amplitude in the very 
forward direction.

Note that a QCD interpretation of the scaling property, cf.  Eq.\eqref{crossscalingzero}, has been 
recently proposed \cite{Peschanski:2024tlr}.

\subsection{2. Scaling amplitudes with positive signature (Pomeron)}
\label{Scaling amplitudes in the Regge formalism}

The problem we want to study is how to describe a scaling 
amplitude such as
Eq.\eqref{amplifit} in the S-matrix formalism of tbe Regge theory 
\cite{chew} without a priori referring to any specific phenomenological Regge model. As is well-known, this is related to analyticity properties of t-channel amplitudes. In this framework, the amplitude ${\CA}^{pp}$  can be 
written  as a sum of two components
\ba
   {\CA}^{pp}_\eta(s,t)\ =\ \fr 1{2\pi i}\int_{\C} {s^{l} +\eta\ (-s)^{l} \over \sin \pi l}\
   \ a_\eta (l,t)\  \ ,
\label{regge} 
\ea
where $\eta =\pm$ is called the signature. $a_\eta(l,t)$ are the analytic 
continuations of the partial-wave amplitudes of both $ \pm$ signatures in 
the crossed two-body $t$-channel. The contour $\C$ is curled around the poles,  due to the positive,  
even or odd, integer zeros of $\sin \pi l,$ depending on  $\eta$, since the poles of opposite $(\mp)$ parity are 
canceled by the numerator.
Indeed,  the integrand of \eqref{regge} can also be written
\eq
\!\!\! {s^{l} +\eta\ (-s)^{l} \over \sin \pi l}
&=& \ \fr {(e^{-i {\pi\over 2}} s)^l}{\sin (\pi l/2)}\quad {\rm if}\ \eta=+1 \n
&=& \ i\ \fr {(e^{-i {\pi\over 2}} s)^l}{\cos (\pi l/2)}\quad {\rm if}\ \eta=-1\ .
\label{reggephase}
\eqx
As shown by Eq.\eqref{reggephase}, the S-matrix Regge formalism \cite{chew} imposes a thorough connection between the $s$ dependence 
and the accompanying complex phases of the amplitudes through the substitution  
everywhere
\ba
s\ \to\ \sg\  =\ e{^{-i {\pi}/ 2}}s \ .
\label{relation}
\ea
The expression, Eq.\eqref{regge}, can be rewritten as,
\eq
\ \ \ \ \ {\CA}^{pp}_+(s,t)\ &=&\ \fr 1{2i\pi}\int_{\C}\!\!dl\ \sg^l\  \hat a_+(l,t)\quad {\rm if}\ \eta=+1\ , \n
\ \ \ \ \ {\CA}^{pp}_-(s,t)\ &=&\ \fr 1{2i\pi}\int_{\C}\!\!dl\ i \sg^l\  \hat a_-(l,t)\quad {\rm if}\ \eta=-1\ , 
\label{reggesimple} 
\eqx
with the redefinitions $\hat a_+(l,t) =  a_+(l,t) /\sin(\pi l/2)$ and $ \hat a_-(l,t)= \ a_-(l,t) /\cos(\pi l/2) \ .$

There are important consequences of 
Eqs.\eqref{reggesimple} since the unique energy dependence of the 
amplitudes $ {\CA}^{pp}_\eta$ is in terms of $\sg.$ Hence, there exists
a precise energy-phase relationship, Eq.\eqref{relation}, predicted by the Regge formalism independently of any phenomenological model.
Moreover,  for a negative  signature amplitude there appears  a phase shift of $\pi/2$ w.r.t. the positive signature amplitude, cf. the $i$ factor in Eq.\eqref{reggephase} for $ \eta = -1$. However the partial wave amplitudes, being a priori complex functions, may have their own phases. 

In the following, we will focus on the positive sign signature which corresponds to the so-called Pomeron t-channel exchange. We shall examine whether a scaling Pomeron amplitude may describe the data as well at least than the good fit obtained with the amplitudes, Eqs.\eqref{amplifit}.
 
Using the complex variable $\sg$, Eq.\eqref{relation}, and  the substitution, Eq.\eqref{relation}, acting on the amplitudes, Eqs.\eqref{amplifit}, we are led to write the scaling Pomeron  amplitudes as 
\eq
{\cal A}^{pp} &=& \ {\cal \tilde A}_{1}-e^{i\tilde\varphi} { \cal \tilde A}_{2}\n
{\cal \tilde A}_{i=1,2}\ &=& \ 
{\tilde N}_i^0 \ 
\left({\sg \over {\rm TeV^2}}\right)^{ \al /2 +1} \!\!
 e^{{\tilde B_i^0}\tau}\ 
\label{amplifitregge}
\eqx
where the tilded amplitudes and parameters are to be determined by a best fit to TOTEM data.
One indeed finds a comparably good fit with such parameterizations \cite{experts}, see Fig.1.
\begin{figure}
\scalebox{0.50}{\includegraphics*{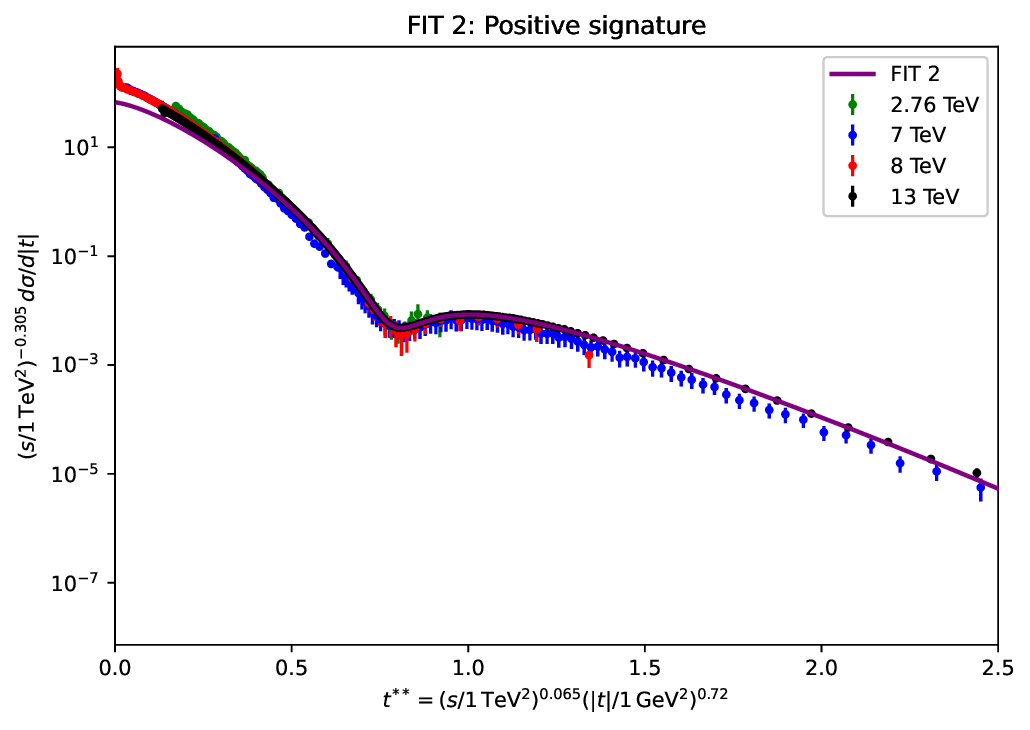}}
\caption{Scaling Pomeron vs. TOTEM data (from \cite{experts}). Both the fit using Eq.\eqref{amplifitregge} and TOTEM data are displayed on the scaling plot  $\log (s^{-\al} d\sigma/dt)\ vs.\ t^{**}$.}
\label{scalingcomponents}
\end{figure}

\subsection{3. Regge formalism: Determination of the t-channel partial waves}
\label{5}
As discussed in  Section 2, the integrand of the Regge 
integral $a(l,t)$ in the first of  Eqs.\eqref
{reggesimple} represents the analytic continuation of the 
partial wave amplitude in the complex $l$-plane. Let us determine 
such analytic continuation for the scaling amplitudes, Eqs.\eqref
{amplifit}.

As a starting point, we make the scaling ansatz. The scaling amplitudes expand  as follows, 
\eq
   {\cal \tilde A}_{i=1,2} &=& 
\ - {\tilde N}_i^0 \ \left({\sg \over { {\rm TeV^2}}}\right)^
{ \al / 2 + 1}
   \ e^{{\tilde B_i^0}t\ \left( {\sg \over{ {\rm TeV^2}}}\right)^{\gm / 2}} \n
 &=&\  -\ {\tilde N}_i^0  \ 
   \sum_{n=0}^\infty {({\tilde B_i^0}t)^n \over n!}\ \left({\sg\over
     { {\rm TeV^2}}}\right)^{1 + \al/ 2  +n  \gm /2}\ .
\label{amplifitreggeexpand}
\eqx
The summation, Eq.\eqref
{amplifitreggeexpand}, can be realized through a sum of residues of poles of the following integrals in the complex $l$-plane
\eq
{\cal \tilde A}_{i=1,2} &=& \ {2i\pi} {{\tilde N}_i^0} \sum_{n=0}^\infty  {({\tilde B_i^0}t)^n \over n!} \int_{\cal C}
  \fr {\sg ^l dl} {l-1-\!\fr {\al +n\gm}2}\ ,
\label{amplifitreggeintegrand}
\eqx
where
the contour ${\cal C} $ goes
around all poles of the integrand. By denoting
\eq
x \equiv {\tilde B_i^0}t\ ,\quad \lm \equiv \fr 2\gm( l-1-\fr \al 2)
\ ,
\label{variables}
\eqx
the sum in Eq.\eqref{amplifitreggeintegrand} can be cast into the two 
equivalent forms  (see  the Appendix).
\eq
 \sum_{n=0}^\infty \ {x^n \over n!}\ 
\fr 1 {\lm-n}\ =  -\ e^x\ \sum_{k=0}^\infty {(-x)^k \ \Gamma(-\lm)\over \Gamma (-\lm+k+1)}\ .
\label{secondform}
\eqx
 Interestingly, the second expression of Eq.\eqref{secondform} is known   \cite{wiki} to be related to the upper incomplete Gamma function $G (u,y)$, 
\eq
e^x\ \sum_{k=0}^\infty {(-x)^k \ \Gamma(-\lm)\over \Gamma (-\lm+k+1)} = \ \Gamma(-\lm)\ G^* (-\lm,-x)\ ,
\label{holomorph}
\eqx
where $G^* (u,y) $ is  the holomorphic factor of 
$G (u,y), $ see the Appendix.

In Eq.\eqref{secondform} only $\Gamma(-\lm)$ 
 exhibits single poles for every  positive integer $\lm$. The scaling ansatz has induced that no other 
 singularity is present with respect to both complex variables. Hence, 
in their   scaling form,  Eq.\eqref
{amplifitregge}, the amplitudes have only 
single pole singularities at positive integer values of $\lm .$ Singularities at infinity in the complex plane may exist but do not seem to be necessary to our analysis.
There are no other Regge singularities in the complex $l$-plane, for instance standard Regge poles or cuts. The scaling form has thus a specific singularity structure. Note however that scaling is only seen at moderate but nonzero mementum transfer, so the full amplitude will probably generate the singularities. The problem remains open to see the connection with the scaling properties.

Since $ \lm$ is a positive integer, it is interesting to replace $ l$ by its value in terms of $ \lm$ in the complex integral 
Eq.\eqref{amplifitreggeintegrand}. Using the identity \eqref
{secondform}, one finally obtains
\eq
{\cal \tilde A}_{1, 2} &=&  \fr {{\tilde N}_i^0} {2i\pi}\ \sg^{1+\al/2}\ \times \n
&\times& \int_{\cal C}\!\ \sg^{\gm\lm /2}\ d\lm\ \Gamma(-\lm)\ 
  G^* (-\lm,-{\tilde B_i^0}t)\ .
\label{amplifitreggeintegrandlambda}
\eqx
Eq.\eqref{amplifitreggeintegrandlambda} provides the complete 
Regge form of the $pp$ elastic amplitude in its scaling form, 
Eq.\eqref{amplifitregge}.

It is interesting to note that the second line of Eq.\eqref
{amplifitreggeintegrandlambda} compares well with the generic 
Regge formula, Eq.\eqref{regge}, if we perform the replacements
\eq
l \to 2\lm\ , \quad \sg \to \sg ^{\gm/2}  ,\ {a}(l,t) \to 
\gm^* (-\lm,-{\tilde B_i^0}t)\ .
\label{replacements}
\eqx

\subsection{Highlights}
In the present paper, we essentially derive two results inspired by the use of scaling amplitudes \cite{scaling} to describe $pp$ elastic data at the LHC. 

$\bullet$ Using the Regge formalism, we define and derive a scaling amplitude with a positive signature, namely the Pomeron. It successfully describes the TOTEM data.\cite{TOTEM}.

$\bullet$ Looking for a t-channel point of view on the scaling amplitudes, we
find the following analytical result, namely an everywhere analytic continuation
of  partial waves of t-channel index $l_t=\lambda$  in the whole $\lm$ 
complex plane up to a rescaling of the complexified  variable $ \sg 
\to \sg ^{\gm/2}.$  This means  a rescaling of the"momentum transfer value in the
t-channel" (i.e. the energy value in the s-channel). This is realized thanks to
the analytic functions $G^* (-\lm,-{\tilde B_i^0}t)$, see the Appendix.

\subsection{Appendix: Proof of relation \eqref{secondform}}
\label{8}
In order to prove Eq.\eqref{secondform}, let us multiply each of the 
two sums by a factor $- x^{-\lm}.$ We then show that both functions 
have equal derivativewith respect to $ x$ and equal values at $x=0, $ for 
all complex values of $\lm.$

\noindent For the left side of  Eq.\eqref{secondform}, the derivative 
gives
\eq
{d \over dx}\left\{- x^{-\lm}\sum_{n=0}^\infty \ {x^n \over n!}\ 
\fr 1 {\lm-n}\right\} &=&
 \sum_{n=0}^\infty  {x^{n-\lm-1} \over n!} \n 
& &= \ x^{-\lm-1}\ e^{x}\ .
\label{left}
\eqx
For the right side of  Eq.\eqref{secondform}, the derivative gives
\eq
{d \over dx}
\left\{
x^{-\lm} e^x\ \Gamma(-\lm)
\sum_{k=0}^\infty {(-x)^k \ \over \Gamma (-\lm+k+1)}
\right\} &=&\n
=\ {d \over dx}
\left\{\ (-1)^{\lm}\  { G}(-\lm,-x) \right\}
= \ x^{-\lm-1}\ e^{x}\ ,
\label{right}
\eqx
where ${G}(u,y) = \int_0^y t^{u-1}e^{-t}$ is the lower 
incomplete gamma function.
Note that the function
\eq
G^*(u,y)\ \equiv \ e^y\ \Gamma(u)
\sum_{k=0}^\infty {y^k \ \over \Gamma (u+k+1)}
\label{gammastar}
\eqx
is known \cite{wiki} to be holomorphic in the complex $\C\otimes\C$ 
space of its $u,y$ variables.

\subsection{Acknowledgements}
We thank C.~Baldonegro, J.~Corral and Ch.~Royon for sharing with us their knowledge of the scaling properties of TOTEM data and  the fitting procedures using scaling amplitudes.

\end{document}